# Interplay of bias-driven charging and the vibrational Stark effect in molecular junctions

Yajing Li[1], Pavlo Zolotavin[1], Peter Doak[2], Leeor Kronik[3], Jeffrey B. Neaton[4,5,6], Douglas Natelson[1,7,8,*]

[1]*Department of Physics and Astronomy, MS 61, Rice University, 6100 Main St., Houston, Texas 77005 USA*

[2]*Center for Nanophase Materials Sciences, Oak Ridge National Laboratory, Oak Ridge, Tennessee 37831-6493 USA*

[3]*Department of Materials and Interfaces, Weizmann Institute of Science, Rehovoth 76100, Israel*

[4]*Molecular Foundry, Lawrence Berkeley National Laboratory, Berkeley, CA 94720 USA*

[5]*Department of Physics, University of California at Berkeley, Berkeley, CA 94720 USA*

[6]*Kavli Energy Nanosciences Institute at Berkeley, Berkeley, CA 94720 USA*

[7]*Department of Electrical and Computer Engineering, MS 366, Rice University, Houston, TX 77005 USA*

[8]*Department of Materials Science and Nanoengineering, MS 325, Rice University, Houston, TX 77005 USA*

*To whom correspondence should be addressed. Emails: natelson@rice.edu

## ABSTRACT

We observe large, reversible, bias driven changes in the vibrational energies of PCBM, based on simultaneous transport and surface-enhanced Raman spectroscopy (SERS) measurements on PCBM-gold junctions. A combination of linear and quadratic shifts in vibrational energies with voltage is analyzed and compared with similar measurements involving $C_{60}$-gold junctions. A theoretical model based on density functional theory (DFT) calculations suggests that both a vibrational Stark effect and bias-induced charging of the junction contribute to the shifts in vibrational energies. In the PCBM case, a linear vibrational Stark effect is observed due to the permanent electric dipole moment of PCBM. The vibrational Stark shifts shown here for PCBM junctions are comparable to or larger than the charging effects that dominate in $C_{60}$ junctions.

An applied voltage across a molecular junction can influence the mechanical coupling between the constituent atoms both by Stark physics (rearrangement of the charge density within the molecule by bias-driven electric fields) and through charge transfer between the molecule and metal electrodes. At the same time, coupling between the electrons and vibrational modes is a critical mechanism for energy transfer in electronic conductors. Probing the relative effects of local electric field and charge state on molecular vibrations therefore lays the groundwork for better understanding of energy dissipation at the nanoscale. Vibrational Stark spectroscopy is one means to investigate the influence of electric field on the dynamics or populations of species undergoing the chemical reactions. Quantitative analysis of the spectral Stark shifts can reveal rich information on variations in the local electric field, and its effect on mutations in biomolecules, conformational changes, and ligand binding[1–4]. The sensitivity of the vibrational transitions to an electric field can also provide a probe into the local electrostatics of an ordered system[5,6].

Previous vibrational Stark effect work has largely focused on analysis of the line-shape evolution of Stark spectra for large ensembles of molecules[7,8]. Methods that provide averaged information may not be sufficient to study surface chemical reactions, because often molecules adsorbed at specific interfacial sites govern surface reactivity.

Surface enhanced Raman spectroscopy (SERS)[9,10] with single-molecule sensitivity[11–13] can be utilized to probe the interfacial electric field in diffuse layers[6] and to study the potential-dependent vibrational frequencies of adsorbates on a variety of transition metal surfaces[14]. The vibrational Stark effect was previously utilized as a measurement tool to infer the plasmonic field enhancement in metallic nanostructures[15–18]; complementary studies, where an applied field is instead used to examine chemical bonding and electrostatic field effects on vibrational modes at the single molecule level, would be of interest and remain challenging. Molecular-scale junctions have proven to be valuable tools for studying vibrational physics[19,20]; as SERS hotspots under conditions of electronic bias they are an enabling technology[21,22] for such studies.

In this article, we report the voltage bias-driven vibrational energy shifts of junctions nominally containing individual PCBM (phenyl-$C_{61}$-butyric acid methyl ester) or $C_{60}$ molecules, at a substrate temperature of 80 K. Analyzing the bias dependence of the vibrational peak energies, we find that statistically the PCBM-containing junctions have noticeably larger linear-in-bias contributions to the peak shifts, compared with $C_{60}$-containing junctions. We compare these observations with density functional theory (DFT) calculations to evaluate the relative importance of the vibrational Stark effect and bias-driven charging mechanisms. The calculations qualitatively reproduce the systematic differences observed between PCBM and $C_{60}$ junction bias dependences, and suggest that while the dominant quadratic shifts in $C_{60}$ junctions are attributable to charging effects[23], the dominant linear shifts in PCBM junctions have significant contributions from vibrational Stark physics. Quantitative discrepancies between the calculation and measurement are attributed to image charge and related electrode effects, explicitly neglected in the theoretical model.

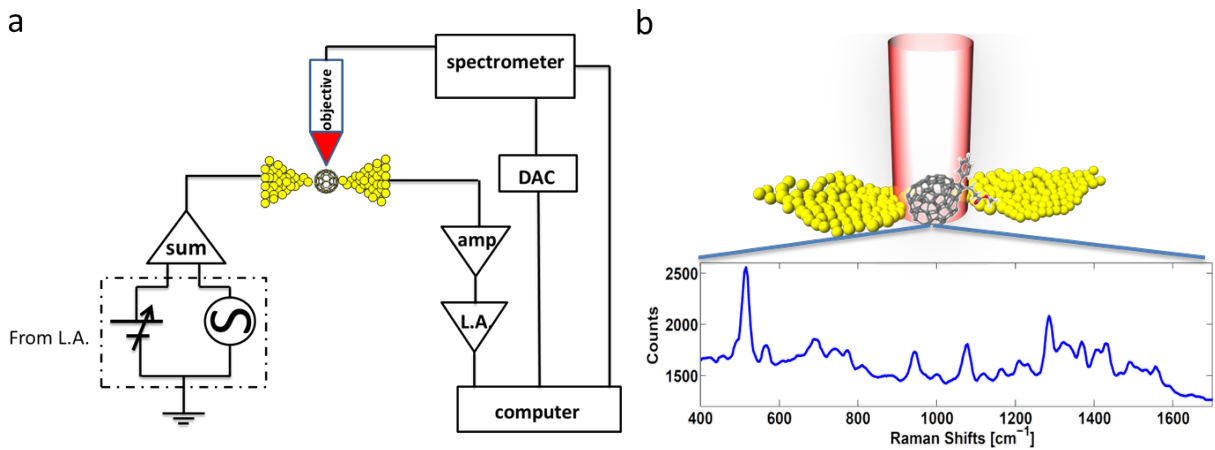

**Figure 1**. (a) Diagram of the Raman measurement setup. (b) Raman spectra of PCBM in an electromigrated junction.

Figure 1a shows the experimental design of the combined Raman spectroscopy and electronic transport measurement system. As described previously [21,22], Au electrodes connected by a nanowire constriction 120 nm wide and several hundred nm in length are fabricated using electron beam lithography on an oxidized Si substrate. The resulting bow-tie structures are cleaned via exposure to oxygen plasma and spin-coated at 1800 rpm with 0.1 mM solution of PCBM or $C_{60}$ in toluene. The devices are wire-bonded to a chip carrier for electrical measurement and are placed in a microscope flow cryostat. The substrate is cooled in high vacuum to 80 K. For each device, following electromigration[24] at this temperature, the constriction is broken to form a tunnel junction with a closest inter-electrode separation on the nanometer scale. The nanometer gap supports localized surface plasmon resonances with large electric field enhancements, sufficient for SERS studies of single molecules[11,12,25]. After the molecular junction is prepared by electromigration, AC and DC biases are applied to the metal electrodes through a summing amplifier to measure the differential conductance at each DC bias. The rms AC bias is typically 10 mV or less, while the DC bias ranges from -0.5 V to 0.5 V, limited by the device stability. A simultaneous Raman measurement is performed using a home-built Raman microscope with a 785 nm diode laser illumination source. Figure 1b shows the Raman spectrum of a typical device. The sharp peak at 520 cm$^{-1}$ originates from the Si substrate; mapping of this Si emission is used to locate the center of the junction. Other peaks shown are believed to be PCBM vibrations. We observe a larger number of PCBM vibrational modes than previously reported [26,27]. One possible explanation for this and similar observations in $C_{60}$ containing junctions[23] is that adsorption in the junction results in a polarizability tensor of lower symmetry than that of the isolated molecule, as the presence of the electrode and nature of the molecule-electrode interaction lifts mode degeneracies and alters selection rules. Multiple molecules in the SERS hotspot is another possibility, though the observation that changes in Raman emission correlate with changes in the (extremely spatially localized[28]) inter-electrode tunneling conductance limits this possibility[29]. Another explanation for the proliferation of modes could be chemical damage due to direct, catalytic, or hot electron photochemistry at the metal interface. Further investigations are ongoing.

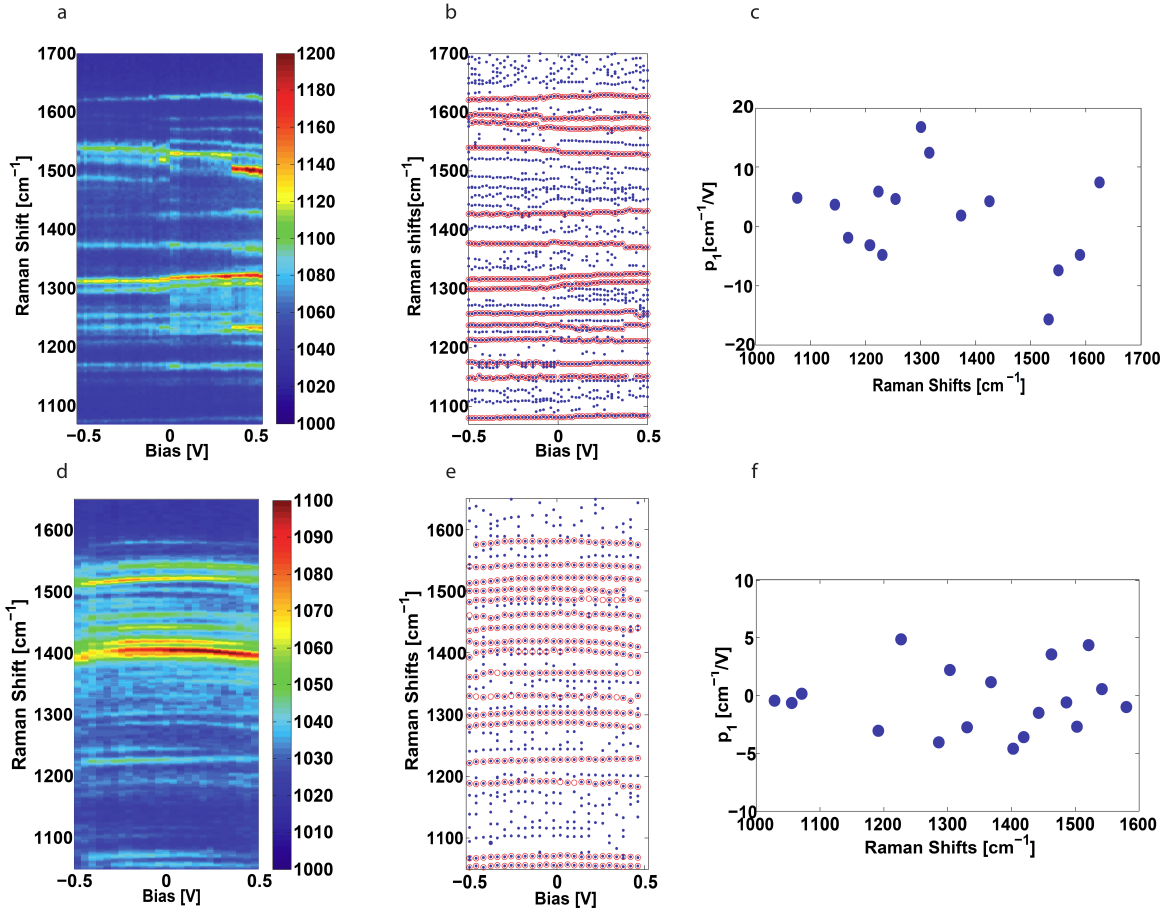

**Figure 2**. Evolution of vibrational modes with applied bias. (a) Stokes Raman emission of PCBM in a typical electromigrated junction as a function of bias. (b) An automated peak-finding

routine generates a map of the evolution of the peak positions of dominant modes in (a), highlighted in pink. (c) Linear bias-dependent tuning rates of each highlighted mode are extracted from fitting the peak position to applied potential according to equation (1). (d) to (f) Corresponding analogous data sets measured on a typical $C_{60}$-containing junction.

Figure 2a shows the Stokes spectra as a function of applied bias for a typical PCBM containing junction. Note that a "blinking" event occurs at 0 V and another at ~ 0.3 V, as evident by the intensity changes of the spectrum. These are likely due to a stochastic configuration change such as reorientation of the molecule with respect to the metal electrodes. For each individual spectrum the peak positions are determined using an automated procedure, and results are shown in Figure 2b. The dominant peaks that display continuous evolution of the mode position with respect to the bias are highlighted in pink. Each vibrational mode has a unique shift as a function of DC bias voltage. Some peak positions do not shift discernably with applied bias, while others increase or decrease by as much as 20 cm$^{-1}$ across the bias window. The energy shifts are fitted to a quadratic function, $v=v_0+p_1V+p_2V^2$. Here $V$ is the applied DC bias, and $p_1$ and $p_2$ are fitting coefficients. For all curves highlighted in Figure 2b, the $|p_2 \times V_{max}|$ is found to be significantly smaller in magnitude than $p_1$, implying that the dominant bias-driven effect is a linear-in-bias shift of the vibrational energy. Figure 2c shows $p_1$ for each vibrational mode highlighted in Figure 2b.

Corresponding equivalent measurements of a representative $C_{60}$-containing junction are shown in Figures 2d, 2e and 2f. There are noticeable differences in the bias dependence of the vibrational modes as compared to PCBM. Of the modes found to shift, the majority shift quadratically in bias. The coefficient of the linear shifts, $p_1$, in the $C_{60}$ junctions is generally smaller than in the PCBM case. The largest magnitude of $p_1$ for this $C_{60}$ junction does not exceed 5 cm$^{-1}$/V. Note that there is some "noise floor" in our ability to determine $p_1$ through peak tracking. The nonzero values of $p_1$ in this $C_{60}$ device show the limits on such an analysis due to the resolution of the spectrometer, precision of peak finding, and spectral blinking. The inherent asymmetry of the junction geometry (e.g., slight differences in work function between source and drain electrodes due to crystallographic asymmetries) can in principle lead to a "built-in" potential at the junction even when the macroscopic applied bias is zero[30]. This would also impose a systematic voltage asymmetry, though this would be expected to affect all peaks equally[23]. Despite the junction-to-junction variation in the Raman spectrum, the sign, the quadratic form, and the magnitude have been consistent over 9 measured PCBM devices; the remaining 7 devices suffer strong stochastic intensity fluctuations and spectral diffusion during the timescale of the measurements, preventing a clear evaluation of the bias-driven shifts in those junctions. As analyzed in previous work[23], the quadratic bias dependence of the $C_{60}$ junctions is believed to originate not from Stark physics, but from the effect of voltage on charge transfer[31,32] between the Au electrodes and the $C_{60}$, together with the dependence of vibrational frequencies on the effective occupation of antibonding molecular orbitals.

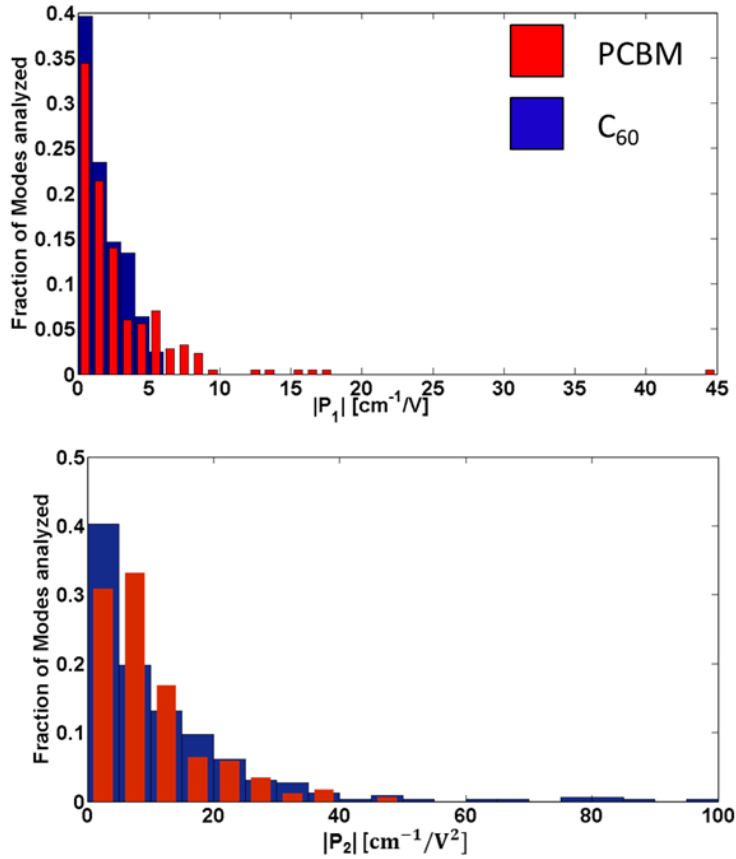

**Figure 3**. Statistical analysis of first order and second order vibrational shifts of PCBM and $C_{60}$. 9 $C_{60}$-containing junctions and 9 PCBM-containing junctions which showed the bias dependence are analyzed. Each junction typically exhibited 20 to 30 modes clearly identified through peak-tracking for analysis according to Eq. (1). Top and bottom panels: normalized histogram of $|p_1|$ and $|p_2|$ distribution, respectively. PCBM data are plotted in red, narrow columns and $C_{60}$ data in blue, wide columns.

The extracted distributions of $|p_1|$ and $|p_2|$ coefficients of modes tracked in the stable PCBM and $C_{60}$ junctions are presented as normalized histograms in Fig 3. For both types of junctions, $p_1$ is smaller than 2 cm$^{-1}$/V for more than half of the modes. Because of stochastic intensity fluctuations, spectral diffusion, and systematics associated with the peak identification and tracking, this represents essentially the lower bound of our ability to identify linear bias dependences of mode energies. We also note that inherent asymmetries in the junction would systematically shift the "zero" of voltage and would cause the fitting procedure of Eq. (1) to find some small linear shift even in the absence of other physics. For C$_{60}$ junctions, the percentage of modes with larger linear shifts decreases rapidly and the distribution cuts off at $p_1$=6 cm$^{-1}$/V. In contrast, for PCBM junctions, the $p_1$ distribution extends much further, to about 20 cm$^{-1}$/V, with one extreme case of $p_1 \approx 42$ cm$^{-1}$/V. The comparatively broad distribution of $p_1$ values of PCBM junctions suggests different mechanisms that are less relevant to C$_{60}$ junctions. We note that this systematic difference between C$_{60}$ and PCBM-containing junctions in the magnitude of linear bias dependences of mode energies is something that the human eye picks out relatively readily from the color plots like Fig. 2a, even without formal quantitative peak-tracking analysis.

The distribution of $p_2$ represents the second order vibrational shifts observed as a function of applied potential. For C$_{60}$, DFT calculations[23] based on imposing an external DC electric field on the molecule in vacuum have shown that shifts caused by the vibrational Stark effect are not systematically quadratic in bias, nor do they favor mode softening. In that work, bias-driven charging of C$_{60}$ in the junction was found to give rise to quadratic-in-bias mode softening of a magnitude comparable to that observed in the experiments. Similar mode softening is observed here, via negative values of $p_2$. For over 60 percent of the modes for both C$_{60}$ and PCBM, $p_2$ is less than 10 cm$^{-1}$/V$^2$, with the distribution decreasing as $p_2$ increases. The $p_2$ distribution of C$_{60}$ is comparatively extended above 20 cm$^{-1}$/V$^2$ and the tail of the distribution goes up to about 100 cm$^{-1}$/V$^2$. This is consistent with charge transfer having a dominant influence on the second order vibrational energy shifts of $C_{60}$. For PCBM, the distribution cuts off at 50 cm$^{-1}$/V$^2$. The bias-driven charging model predicts greater shifts as the energy difference, $E_0$, between the lowest unoccupied molecular orbital (LUMO) and the Au Fermi level is decreased. The difference in $p_2$ distribution between PCBM and $C_{60}$ may indicate a species-specific difference in molecular level alignment and/or a different coupling between the molecule and electrodes.

To understand the so-far unique linear-in-bias Raman shift behavior of PCBM junctions, we use DFT to compute the vibrational frequencies of PCBM as a function of both external electric field and partial charge. As discussed above, prior work on $C_{60}$ junctions has shown that bias-driven changes of the $C_{60}$ charge state can lead to significant negative shifts in vibrational mode frequencies but with no significant linear dependence of the shift on bias. PCBM, despite being a fullerene derivative, is quite asymmetric, and because of the presence of a permanent dipole moment and large polarizability we would expect a significant linear vibrational Stark effect with bias, in addition to any charging effects. To explore this quantitatively for PCBM junctions, we neglect explicit treatment of the electrodes and instead model the PCBM junction as a function of bias with a gas-phase PCBM molecule in an external electric field, with steady-state charge derived from a single-level Lorentzian model and assuming coherent tunneling, as was done previously[23] for $C_{60}$ (details of the calculation are given in the supporting information, SI). We then fit the calculated bias dependence of the mode energies to predict the parameters $p_1$ and $p_2$ for each mode.

Importantly, experimental electron affinities (EA) for gas-phase PCBM and $C_{60}$ are quite similar: 2.63 eV and 2.68 eV, respectively[33]. This agrees well with our extended basis set DFT calculations, which yielded EAs of 2.8 eV and 2.9 eV, respectively.[*] Furthermore, in both molecules the LUMO interacting with the Fermi level electrons of the Au contact has a very similar character in extent and symmetry. Therefore, if the PCBM "tail" does not interfere with bonding it is reasonable to expect the Au Fermi level to LUMO energy difference, $E_0$, and the Lorentzian broadening, $\Gamma$, for both molecules to be quite similar. Therefore we use the same values - $E_0$ =0.8 eV and $\Gamma$=0.10 eV – for both molecules. This case would lead to approximately maximal charge transfer (smallest $E_0$ and largest $\Gamma$) for PCBM.

In Fig. 4, we plot the calculated $p_1$ and $p_2$ values derived from our model for each mode for both $C_{60}$ and PCBM.[†] For the above model parameters the calculated bias-induced vibrational shifts for PCBM, including both field and charging effects, show linear components of up to ~ 2 cm$^{-1}$/V and quadratic elements of up to ~7 cm$^{-1}$/V$^2$ in magnitude, at maximum field (1.4 V/nm). In contrast, for our calculations of $C_{60}$ using the same parameters, linear shifts with bias show a maximum of ~0.4 cm$^{-1}$/V and quadratic shifts of up to 20 cm$^{-1}$/V$^2$.

---

[*] Importantly, standard basis set calculations resulted in *much* lower electron affinity values, of 1.53 eV and 1.51 eV for PCBM and $C_{60}$, respectively. However, sampling of a variety of field and charge conditions showed that extended and standard basis sets produce similar results for the vibrational shifts. Therefore the results shown here were all computed with standard basis sets – see the SI for more details

[†] Importantly, the calculated vibrational spectrum for gas-phase PCBM is in good agreement with prior work. For example, for prominent modes such as the primary C-H$_2$/C-O peak and C=O stretch found at 1163 cm$^{-1}$ and 1745 cm$^{-1}$, respectively, our calculations agree well with previous calculations[34] (1152 cm$^{-1}$ and 1732 cm$^{-1}$) and experiments[35,36] (1187 and 1738, 1740 cm$^{-1}$).]

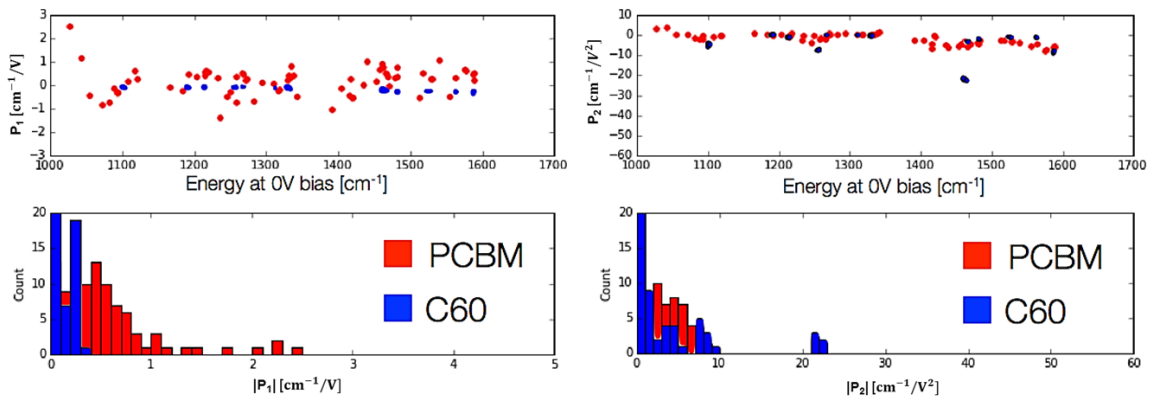

**Figure 4**. Fits to the expression $\nu=\nu_0+p_1V+p_2V^2$ for data computed with our theoretical model, including bias-induced charging and field (see SI for details). (a) Top plot: $p_1$, the linear fit coefficient of shift with respect to the zero bias normal mode energy. Bottom plot: a histogram of the absolute values of all shifts, obtained with 0.1 cm$^{-1}$/V binning. (b) Top plot: $p_2$, the quadratic fit coefficient of shift with respect to the zero bias normal mode energy. Bottom plot: a histogram of the absolute values of all shifts, obtained with 1 cm$^{-1}$/V binning.

To lowest order, the linear term in the vibrational Stark effect is known to originate with the second derivative of the induced dipole moment with mode displacement[37], which will only be nonzero for IR active (or polar) vibrational modes. As PCBM lacks inversion symmetry, it features more IR active modes than centrosymmetric C$_{60}$; this is consistent with the greater linear shifts (and $p_1$ values) exhibited by PCBM modes. Future atomistic treatment of both molecules beyond the gas-phase including electrodes would be desirable to further explore the role of the electrodes, the bias (and associated electric field), and charging in more detail.

Binned in analogous fashion, our calculations for each mode of both $p_1$ and $p_2$ are strikingly consistent with the statistical trends in Fig. 3, though the calculated magnitudes are smaller than the experimental values, by factors of ~ 10 and 4, respectively. While small changes in $E_0$ and $\Gamma$ can bring the calculated $p_2$ values for the $C_{60}$ junctions into better quantitative agreement with the experiments, we find that no such adjustments can significantly increase the calculated PCBM $p_1$ coefficient, even allowing model parameters that lead to more complex bias dependences (*e.g.*, small values of $E_0/\Gamma$ such that the metal Fermi level approaches resonance with the LUMO Lorentzian).

Although successful qualitatively, our model necessarily misses some essential physics of the real device structures that could magnify the impact of the PCBM linear-in-bias response. For example, by not treating the full junction environment, including the metal electrodes and their static and dynamic screening[38], the present calculation explicitly neglects, e.g., image charge physics. A classical "toy model" of a particle harmonically bound to a conducting surface that incorporates image charge effects will show, over some range of parameters, an approximately linear-in-bias shift of vibrational frequency (see SI). Further investigations, including more sophisticated calculations that include static and dynamic screening effects in realistic junction geometries, should constrain this possibility.

*Acknowledgments*. YL and DN acknowledge support from Robert A. Welch Foundation grant C-1636. PZ and DN acknowledge support from ARO award W911-NF-13-0476. Work by PD and JBN was supported by the U.S. Department of Energy, Office of Basic Energy Sciences, Materials Sciences and Engineering Division, under Contract No. DE-AC02-05CH11231. Portions of this work at the Molecular Foundry were supported by the Office of Science, Office of Basic Energy Sciences, of the U.S. Department of Energy under the same contract number.

Computational resources provided by NERSC. Work by LK was supported by the Israel Science Foundation and the Lise Meitner Center for Computational Chemistry.

***Competing financial interests***. The authors declare no competing financial interests. Correspondence and requests for materials should be addressed to natelson@rice.edu.